\documentclass[10pt,letterpaper,twocolumn]{article}
\usepackage{cite}
\usepackage{amsmath,amssymb,amsfonts}
\usepackage{algorithmic}
\usepackage{graphicx}
\usepackage{textcomp}
\usepackage{xcolor}
\usepackage{multirow}
\usepackage{url}

\usepackage[a4paper, total={6.9in, 9.2in}]{geometry}
\def\BibTeX{{\rm B\kern-.05em{\sc i\kern-.025em b}\kern-.08em
    T\kern-.1667em\lower.7ex\hbox{E}\kern-.125emX}}

\begin{document}

\title{\huge A 950 MHz SIMT Soft Processor}

\date{}

\author{Martin Langhammer\\
\textit{FPGA Architecture} \\
\textit{Altera Corporation}\\
London, UK \\
martin.langhammer@altera.com
\and
{Gregg Baeckler}\\
\textit{FPGA IPSE} \\
\textit{Altera Corporation}\\
San Jose, USA \\
gregg.baeckler@altera.com
\and
{Kim Bozman}\\
\textit{FPGA Software} \\
\textit{Altera Corporation}\\
Toronto, Canada \\
kimberly.bozman@altera.com}

\maketitle

\begin{abstract}
Although modern FPGAs have a performance potential of a 1 GHz clock frequency - with both clock networks and embedded blocks such as memories and DSP Blocks capable of these clock rates - user implementations approaching this speed are rarely realized in practice. This is especially true of complex designs such as soft processors.

In this work we implement a soft GPGPU which exceeds 950 MHz in an Altera Agilex-7 FPGA. The architecture is a 32-bit fixed point Single Instruction, Multiple Thread (SIMT) design, with parameterized thread and register spaces. Up to 4096 threads and 64K registers can be specified by the user. In one example, a processor with 16K registers and a 16KB shared memory required approximately 7K ALMs, 99 M20K memories, and 32 DSP Blocks.

\end{abstract}

\section{Introduction}

FPGAs are often used as accelerators, leveraging high internal memory bandwidth and DSP density. This utility is enhanced by the flexibility in data movement (for example, the ability to reach multiple destinations simultaneously) which offers significant advantages over other programmable solutions. Although the FPGA is not an ASIC, modern FPGAs contain many embedded components that are ASICs, providing deterministic performance. Many modern FPGAs have a performance potential of 1 GHz, supported by the embedded ASIC blocks and a predefined clock network; however, most realized designs are often much slower. A survey of papers at a recent major FPGA conference~\cite{FPL2024_proceedings} showed that reported average speeds were 275 MHz, less than 30\% of the FPGA capability.

The FPGA is harder to use than a software programmed device, and there are many more software developers than RTL coders. FPGA design iteration is costly and  time consuming. Some algorithms are difficult to program in RTL, but easy in software. A high performance soft processor could greatly enhance the FPGA by bridging a gap between these two worlds, allowing software acceleration and hardware acceleration to co-exist. But existing soft processors are typically low performance single threaded RISC, with a modest speed, typically around 300MHz ~\cite{Nios}~\cite{NiosV}~\cite{Microblaze}. Although there are a number of previously published SIMT processors ~\cite{FlexGrip,MIAOW,FGPU,SCRATCH,DOGPU,Guppy,Kingyens} and vector processors ~\cite{Vegas,Venice,VectorBlox} these also suffer from a low clock rate, and are sometimes very large.

This work is based on the eGPU (\textbf{e}mbedded \textbf{GPU}) project~\cite{ISFPGA_eGPU} which introduced an area-efficient (${\approx}$10K ALMs), high-performance (771 MHz operating frequency) SIMT processor for Altera/Intel Agilex FPGAs.  The eGPU was intended for embedded applications that may be commonly found in FPGA systems, with the ability to act as both an accelerator and a controller ({\em{i.e.}} managing other, more traditional FPGA accelerator cores). The availability of a high-performance signal processing compute engine (the eGPU is designed as a GPGPU, meant for general-purpose algorithms rather than graphics) allows the implementation of complex algorithms that may be very time-consuming to code and/or debug.

We start with the published eGPU architecture, and build a new SIMT processor that is designed to approach the FPGA design limit of 1 GHz. This required significant changes to the instruction fetch/decode/sequencer and the core ALU sections.

We make the following contributions in this work:
\begin{itemize}
    \item Develop and report a parallel SIMT processor running in excess of 950MHz.
    \item Describe methods for approaching the 1 GHz FPGA design limit in the general case, and for processors in particular.
    \item Evaluate multiple fitting approaches, including multi-instance cases, and evaluate tools characteristics on realized performance.
\end{itemize}

\begin{figure*}
    \centering
    \includegraphics[scale=0.60]{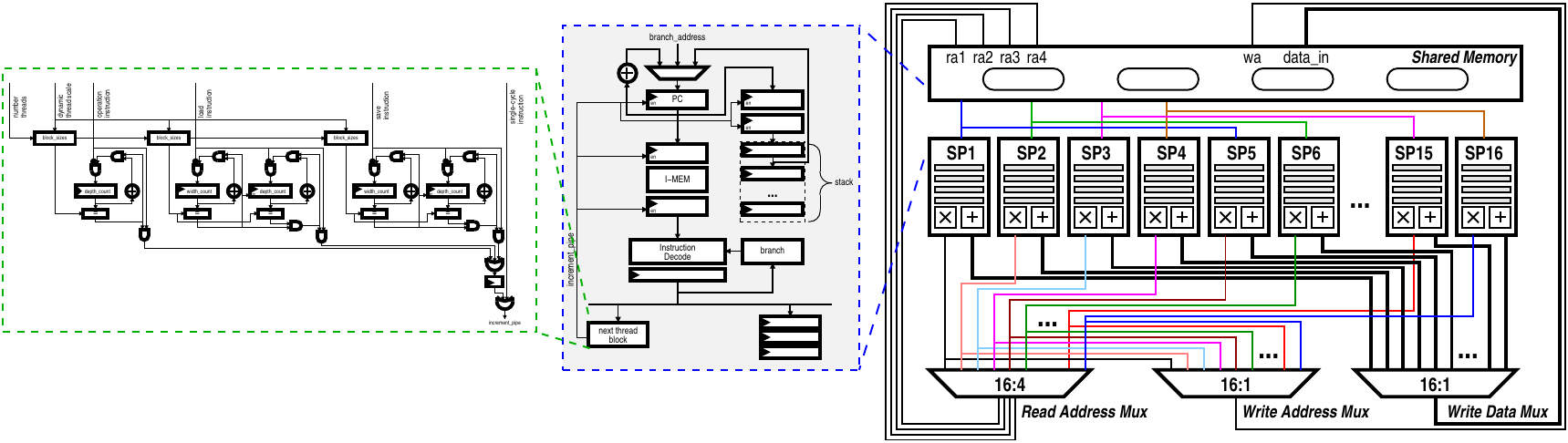}
   \caption{eGPU Top Level Block Diagram}
   \label{fig:top_level}
\end{figure*}

\section{Existing eGPU Architecture}

The eGPU is designed as a single SM (streaming multi-processor) with 16 SPs (scalar processors). The eGPU supports up to 4096 threads, 64K registers, and has a number of user-configurable features, which allow for resource/performance trade-offs.

An instruction fetch/decode/thread sequence block controls the SPs and shared memory. For simplicity, all threads run in lockstep {\em{i.e}} every thread in the current instruction is issued before the next instruction is started.  The instruction set is inspired by Nvidia PTX~\cite{Nvidia_PTX}, with a subset of 61 instructions supported. Predicates (essentially IF/THEN/ELSE for GPUs) are optionally added by a configuration parameter, as they typically increase the logic resources of the processor by 50\%. Predicates are rarely required for many of the embedded application programs.

The processor is comprised of 16 SPs, which was found to provide a good balance between processor performance and the limitations due to shared memory bandwidth. The shared memory architecture is multi-port, a departure from the banked memory typically found in commercial GPGPUs. The multi-port memory (configured as 4R-1W) has a lower potential bandwidth, but a much simpler arbitration mechanism, which is important for saving logic, routing, and latency in the FPGA target. The memory bandwidth reduction is partially offset by dynamic thread scaling, which allows the thread space to be changed on an instruction-by-instruction basis. For example, writing back only a subset of the threads (this may happen during vector reductions), can significantly reduce the number of clocks required for the STO (store) instruction.

Figure~\ref{fig:top_level} shows the structure of the eGPU. In this figure, the instruction fetch and decode block is attached to the left side of the SPs and shared memory. The thread sequencer is called out from this block. Both of these blocks required significant modifications to increase performance from the 750 MHz range to the near GHz range, and are the subject of the next section.

\subsection{Towards a 1 GHz Soft Processor}

The Agilex-7 DSP Block~\cite{Agilex_DSP} configured in floating point mode has a maximum operating frequency of 771 MHz, which in turn limits the performance of the soft SIMT Processor. In order to approach 1 GHz, the architecture must be switched to an integer-only design (the DSP Block runs up to 958 MHz in some of the integer modes). Although this will affect many of the matrix and signal processing applications that may be implemented on the soft GPGPU, integer versions of these have historically been used on fixed-point DSP processors.

\subsection{High Performance FPGA Design Practices}

There are very few published FPGA designs that exceed 800 MHz. Getting a processor, especially a parallel processor, to approach the 1 GHz FPGA capability will be challenging. Understanding how to structure a design and map it to an FPGA will be key.

FPGA logic can be deeply pipelined - there is a register available after each logic function - in the Agilex Adaptive Logic Module (ALM)~\cite{Agilex} the fracturable 6 LUT is combined with four registers. The 6 LUT can be decomposed into two 4 LUTs (or a 2 bit adder segment), and each of the resultant two logic functions can be followed by a register. The two other registers can be independently accessed from outside the ALM to provide a balancing or delay register. Delays can therefore easily be added wherever desired ({\em{i.e.}} independently of a logic function). While delays can easily be inserted, there is a limit as to how much pipelining can speed up a circuit, as extensive logic decomposition (as opposed to merely adding delays before and after the circuit) will eventually affect the placement of the design.

Understanding the FPGA macro-architecture is also important. Agilex devices are comprised of sectors, which encompass a single clock region. Components in the sector have a fixed spatial relationship; ideally the design should be structured to reflect the resources in both count and distances between them. Sectors vary in size, but one representative sector contains 16640 ALMs, 240 M20K memory blocks, and 160 DSP Blocks. 

\section{1GHz Instruction Fetch and Decoder}

All threads in our SIMT processor operate in lockstep. Every instruction, whether a single clock or hundreds of clocks, completes before the next one is issued. This simplifies the instruction fetch and decode architecture, and still enables high performance for the intended embedded applications. 

Although the instruction fetch and decode component is one of the smallest major elements in the processor, it also contains some of the deepest combinatorial logic paths. The performance impact of these paths is mitigated by their width, as they often reduce to a single bit. (Finding a fast path for a single bit is almost always successful). The most complex portion of the instruction decode is the pipeline advance control. This circuit takes into account the cycle times of multiple different types of instructions. In addition, the number of threads is set on a program by program basis, and many of the instructions can control the number of threads run on an instruction by instruction basis. For some instructions, this means that both the thread block width and depth can change, while in others it is simply reflected in the thread block depth.

\begin{figure}
	\begin{center}
		\includegraphics[scale=0.80]{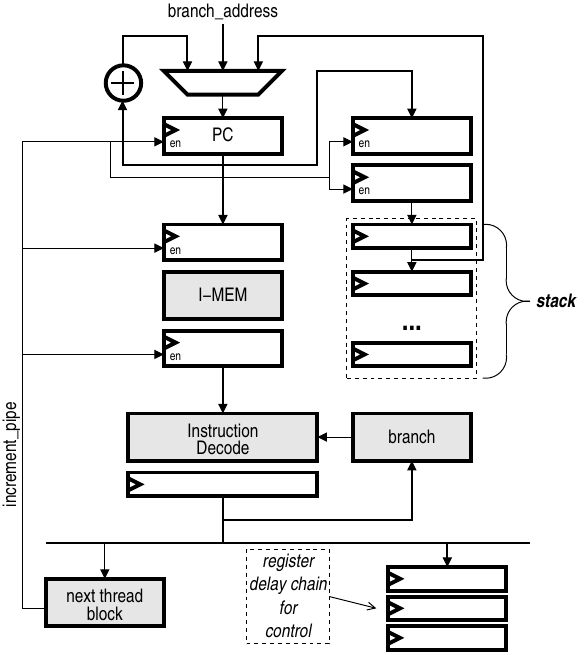}
		\caption{Instruction Fetch and Decode Architecture}
		\label{fig:fetch} 
	\end{center}
\end{figure}

Figure~\ref{fig:fetch} shows the structure of the instruction fetch and decode. This block is deeply pipelined for speed, which requires a short history of addresses to be kept for determining branch returns, and also a mechanism for zeroing already decoded instructions. Some details are omitted for figure clarity, such as the single-cycle DSP processor-like loop instructions. The instruction memory (I-Mem) is also externally re-loadable. A branch taken zeroes out the following instructions in the pipeline.

The decoded control bits and busses to the main core are inserted into a register delay chain. This has a number of advantages, one of which is that this instruction block can be placed independently of the main core. Even if the main core is constrained to one location, the smaller, but more complex instruction block can be placed elsewhere on the device where convenient. As control flow decisions are made entirely in the instruction block, additional pipeline stages can be added to the delay block if needed. This also gives the fitter more flexibility in arranging the placement of the logic inside the instruction block to shorten the pipeline control enable paths, which will likely be the single most critical path in the entire processor. 

\begin{figure*}
    \centering
    \includegraphics[scale=0.65]{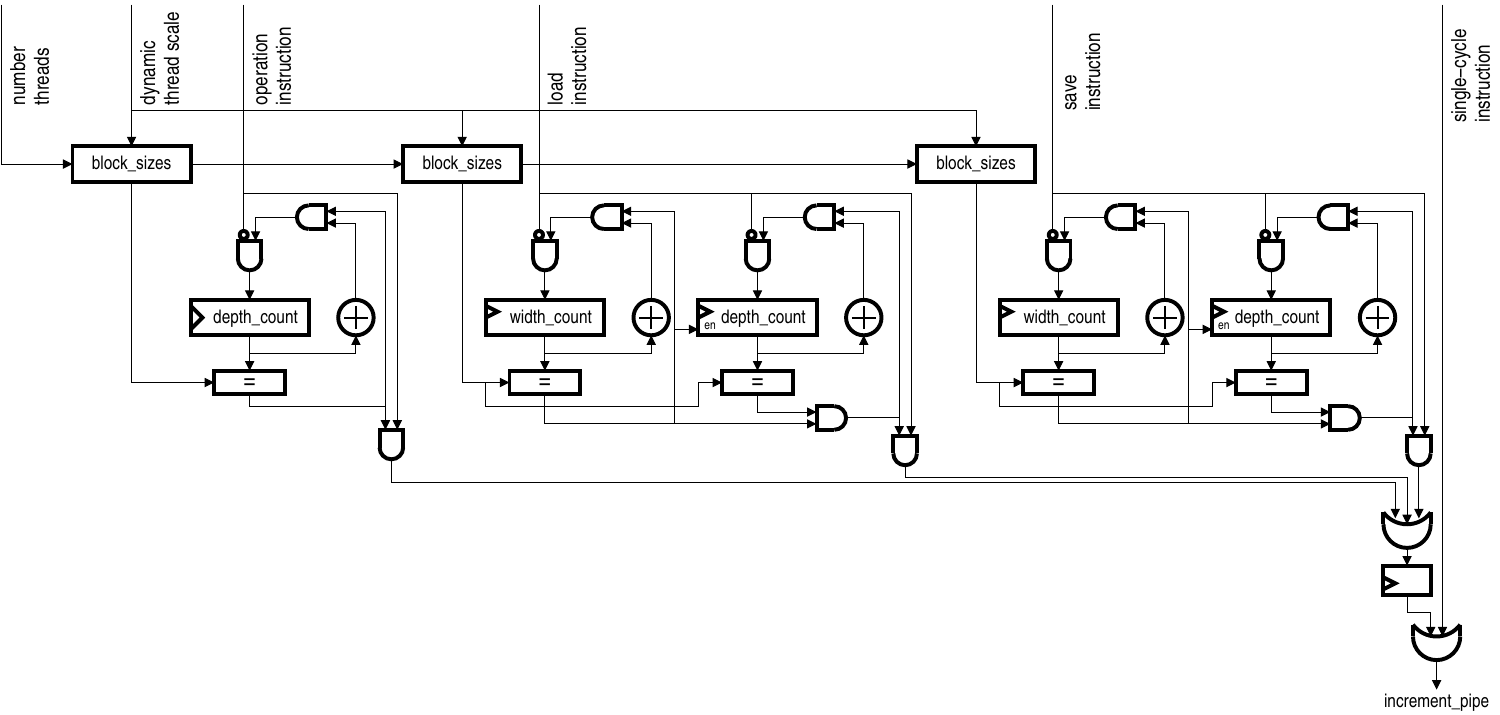}
   \caption{Instruction Fetch and Decode - Pipeline Control}
   \label{fig:next_pc}
\end{figure*}

The pipeline control block - which increments the program counter and advances the pipeline at the end of an instruction - is shown in figure~\ref{fig:next_pc}. The operation instructions ({\em{e.g}} multiply, add, AND, etc.) are counted by thread block depth only, while the load and store instructions are counted both by block depth and width. 

\subsection{Improving Thread Block Performance}

The end of an instruction is defined when the number of clocks that instruction requires has been reached. This signal is now registered to improve performance, so the circuit must check for the number of cycles minus one. In the case of an operation instruction, this is simple, as it is the thread block depth minus 1. For this processor with a parallelism of 16 (the number of SPs), an application example with 512 threads would require 32 clocks (512/16) per operation instruction. The counter would then count 30 cycles (0 to (31-1)). A load instruction would require 4 clocks per block width, and run for a depth of 32. The width counter would count modulo 3, at which point the load depth counter would be incremented. In this case the end of the load instruction would be signalled when the depth was 31, but the width was only at 2, which is the width and depth combination one cycle before the end. The save instruction would be handled in a similar way. 

If dynamic scaling was applied to an instruction, the width and depth count values would be calculated by the associated block size circuit. There is the possibility of an instruction that requires only a single clock cycle, a case which needs separate processing. This case is trapped by the previous instruction decode pipeline stage, and asserts the single-cycle instruction signal (which also detects original single-cycle instructions such as zero-overhead loops).  

\section{High Performance ALU}

\begin{figure}
    \centering
    \includegraphics[scale=0.80]{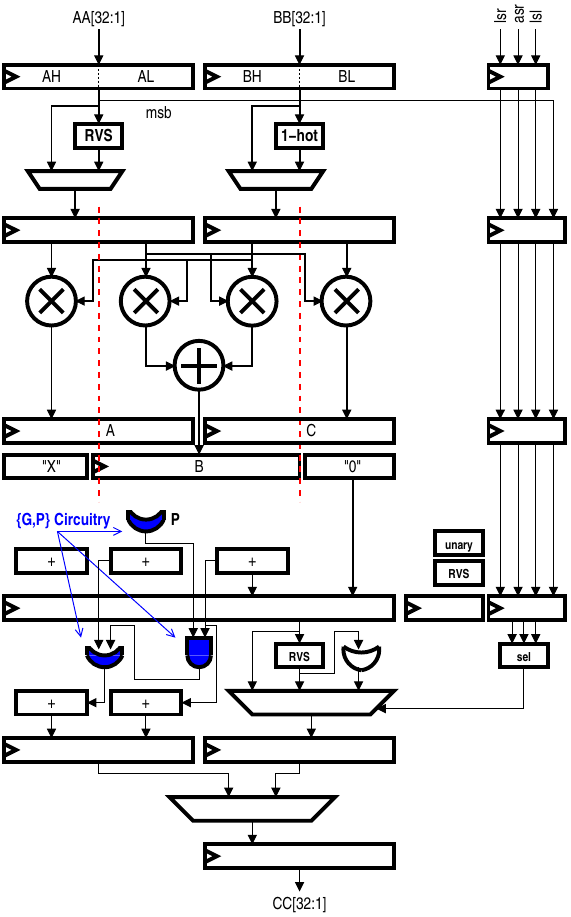}
   \caption{Integer ALU - Including Multiplier and Shifter Units}
   \label{fig:int_alu}
\end{figure}

\begin{figure*}
    \centering
    \includegraphics[scale=0.60]{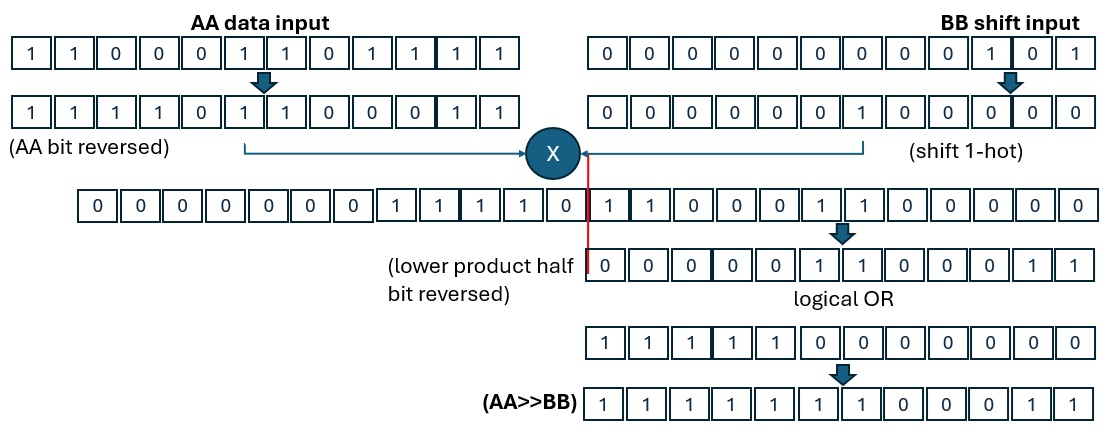}
   \caption{Arithmetic Shift Right: 12-bit example}
   \label{fig:asr}
\end{figure*}

 The integer datapath requires a multiplier; in the most straightforward implementation, we could just use a subset of the Nvidia PTX~\cite{Nvidia_PTX} 24-bit integer multiplier, which is directly supported by the DSP Block with its embedded 27-bit multiplier. For a general solution, however, we implement a 32-bit integer multiplier, with the option of writing back either the high or low 32-bit halves of the 64-bit result (in a GPGPU context, the high value would typically be used for signal processing, and the low value for address generation).

As a 32x32 multiplier is not directly supported in the Agilex devices, it must be constructed from a combination of DSP Blocks and soft logic. The soft logic calculation of the 64-bit output is the  likely bottleneck in the ALU construct. (The DSP Block has a maximum speed of 958 MHz, which will be our target speed for this design. To repeatably and reliably close timing at this clock frequency, the soft logic portion of the design must be comfortably faster than this).

The logic ALU (we will include addition and subtraction as part of this, as they are mapped to soft logic) also needs to meet the 1 GHz target. Soft logic performance can vary depending on many factors - density of design, care of placement, synthesis assignments (which may be needed for other parts of the system) and compile seed values.

The multiplier datapath - despite the constituent DSP Blocks being ASIC constructs - requires a deeper pipeline than the logic datapath. The DSP Block itself has three pipeline stages here: one input and output stage (also needed to provide timing margin to and from the soft logic source and destination), and an internal stage as there are multiple constituent components that need to be combined to support the many configuration options. 

Looking at the required functions, the standard bitwise logic functions (such as AND, OR, XOR) will be able to achieve 1 GHz in a single level of logic. Somewhat more complex bitwise functions, such as cNOT, will likely not be able to reach this performance level in a single level of logic, but as there are a large number of pipeline levels to use (the soft logic ALU is depth matched to the DSP Block datapath) there is considerable flexibility available. The adder function - also supporting operations such as subtraction and absolute value - is implemented as a two stage pipelined adder; the two halves map into a subset of a Logic Array Block (LAB). (The LAB~\cite{Agilex_Logic} is a group of 10 ALMs, which share a common local routing network). The 20-bit adder in the LAB easily meets the 1 GHz performance target. 

A performance problem - which is not evident at the individual function level, but is seen when the entire processor is assembled - is in the shifter. A 32-bit barrel shifter in soft logic is most commonly implemented as a 5-level binary shift. When this structure is pipelined with a single internal register stage, the 1 GHz clock rate is easily achieved, as is when the shifter is included in the soft logic integer ALU and compiled as part of a complete SP. But when multiple SPs are combined to make a SM, we found that the critical path was usually in the shifter, typically reducing the performance below 850 MHz. Tracing the slowest paths showed that the majority of these are in the longer shift levels. For example, a 32-bit, 5-level shifter is comprised of 1-bit, 2-bit, 4-bit, 8-bit, and 16-bit shifts. The 16-bit shifts in particular introduce connections which travel a long way horizontally - {\em{e.g.}} the input to any given ALM in this level will come from two different LABs. The previous combinatorial level also has a long horizontal component (8-bit) to it; it appears that two consecutive logic levels with long routing distances can close timing when compiled as part of a smaller circuit, but placement in a larger system design context is difficult.   

One way to mitigate this is to create a shifter with two internal pipeline stages (such as adding an additional stage between the 8-bit and 16-bit shift levels), but this will also further restrict the flexibility of the placement of the shifters. There is also a resource consideration. A 32-bit shifter requires approximately 50 ALMs, or 100 ALMs for a left and right shift pair. The shifters are a major contributor to the area of the soft processor: the shift pairs in the 16 SPs make up almost 1/4 the total soft logic ($c.$7000 ALMs) of the processor. An alternate circuit that would be more area efficient and provided an easier placement would be desirable here.

\subsection{INT32 Multiplier}

The 32x32 integer multiplier is implemented as a 33x33 signed unit which can support both signed and unsigned numerics. This structure is shown in figure~\ref{fig:int_alu}. The two input operands are split into high and low halves, \{AH,AL\} and \{BH,BL\}, respectively. Four 18x19 multipliers are used over two DSP Blocks, with the input values routed to the 16 LSBs of each multiplier. In the case of an unsigned multiplication the upper bits into all four multipliers are zeroed, otherwise the lower half values are zeroed and the upper half values sign extended.

One of the DSP Blocks is configured with two independent multipliers, which process AH*BH and AL*BL, respectively. This DSP Block outputs the \textbf{A} and \textbf{C} vectors. The other DSP Block is configured to calculate the sum of two multipliers, which processes AH*BL and AL*BH, output as the \textbf{B} vector. The three 37-bit vectors are arranged as follows to make two 66-bit vectors: the lower 34-bits of vector \textbf{A} are appended to the left of the lower 32-bits of vector \textbf{C}, and all 37-bits of vector \textbf{B} have a 16-bit zero vector appended to the right, and are sign-extended to the left.   

The two resultant DSP Block vectors are then added. Building a structure to consistently close timing at 1GHz for a 66-bit integer addition, which is part of a larger design (without resorting to a deeply pipelined circuit) was solved using a prefix structure to compute carry look-aheads. The 16 LSBs of the result are simply the 16 LSBs of \textbf{C}, and do not require any processing. The next 16 bits therefore do not have a carry-in and are the addition of the two DSP Block vectors for that bit range. The next two 16-bit values are added independently, with the carries added in the next pipeline stage. (The lower 32 bits are delayed to align with the calculation of the upper 32 bits, but the ALM logic in this level is not wasted, and used for the implementation of the high speed shifter, which is explained later in this section).

The carries into the two upper 16 bit segments is calculated using a \{generate,propagate\} ({\em{\{g,p\}}}) pair. The propagate value, which is calculated for the third segment ({\em{i.e}} bits [47:32]) can be calculated independently of any other operation in the previous pipeline level, and registered as a single bit. The input carries then require only a single gate each. (A propagate occurs when an input carry would be propagated through all 16 bits of the segment addition, which is is calculated here using a logical AND of the logical OR of every bit pair of the two operands).

\subsection{Integrated Shifter}

The left and right shifters can be incorporated into the multiplier datapath. The \textbf{AA} input is the data value, and the shift value is decoded from \textbf{BB}. The shift value is converted to a one-hot representation ({\em{e.g.} a decimal '5' is "00...001000"}), which is accomplished in a single level of logic. A value greater than decimal 31 is converted to a one-hot value of all zeroes, in other other words the multiplicative shift result is 0. This is the equivalent of having the data value shifted out of range. Left shifts are simply the multiplication of the data value \textbf{AA} and the one-hot shift value. All shift results are output from the lower 32-bits of the multiplier datapath.
Right logical shifts are accomplished by bit reversing both the data value \textbf{AA} and the output of the multiplier.

Right arithmetic shifts are required, especially as this is a fixed point only processor. Integer arithmetic will be used for all algorithmic processing (rather than the floating point in the more typical GPGPU case), so scaling and normalization (to prevent overflow and control wordgrowth) will need arithmetic (where the sign of the 2's complement number is maintained) right shifts. The direct multiplicative right shift described above can only implement a logical right shift (the leading bits will always be zero, {\em{i.e.}} only unsigned numbers can be supported).

We solve this problem by explicitly calculating the leading ones in the case of a negative signed number. The 5-bit shift value (from input \textbf{BB}) is forwarded to the pipeline location aligned with the DSP Block outputs, where it is converted into a unary number. This number is then bit-reversed (a free operation in hardware) and registered. In the case of a right shift where the MSB of the input value is '1', the reversed unary number is ORed with the bit reversed 32 LSBs of the multiplier.

An example of the arithmetic right shift flow (using 12-bit numbers for brevity) is shown in Figure~\ref{fig:asr}. An input "110001101111" (-913 decimal) is right shifted by 5 bits. The input is first bit reversed "111101100011", and the shift value is converted to a one-hot "000000100000". (A shift by zero would result in a one-hot value of "1", and a shift out of range a one-hot value of "0"). The lower half of the multiplier result is bit-reversed again (the upper half result is not used by the shift calculation). The result is an unsigned number, with leading zeroes. The bit-reversed unary equivalent of the shift value (five '1's) is then ORed with the bit reversed multiplier result to produce "11111100011" (-29). (-913$\gg$5 $\approx$-29).

\section{Results}

We ran several compiles - unconstrained and constrained - to validate the performance of the soft processor over a wide range of possible system uses. The unconstrained compile used the Quartus default synthesis and fitting assignments, other than turning {\em{auto-shift-register-replacement}} to OFF. (Replacing discrete registers with an ALM in memory mode is more area efficient, but impacts our processor as the ALM clock rate is {\em{only}} 850 MHz when configured in this mode). 

\begin{figure}
    \centering
    \includegraphics[scale=0.65]{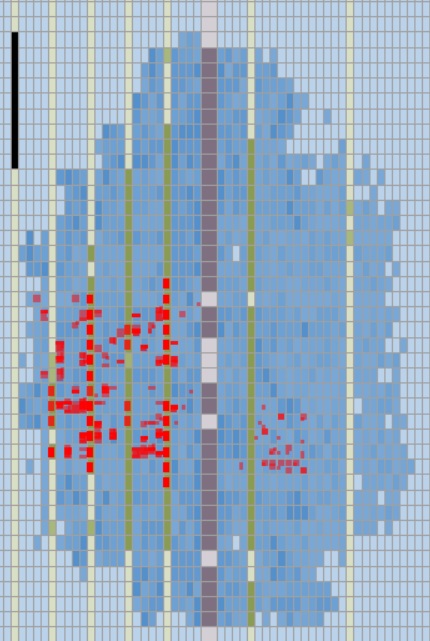}
   \caption{Unconstrained Placement}
   \label{fig:splat_floorplan}
\end{figure}

We used Quartus Prime Pro Edition 24.3 for all our compiles, targeting an Agilex AGFD019R24C21V device. This device contains only one DSP column per sector; as the processor requires two DSP Blocks per SP, placement of the cores is always forced into a 32 row height. Despite the reduced placement freedom, the unconstrained compile achieved 984 MHz, with a restricted Fmax of 956 MHz, which was limited by the DSP Blocks. The placement showed good regularity, creating a near-rectangular layout (see Figure~\ref{fig:splat_floorplan}). The shared memory (highlighted in red) forms a cluster to the left side of the placement, with the 16 SPs straddling the spine of DSP Blocks down the center.

\begin{figure}
    \centering
    \includegraphics[scale=0.60]{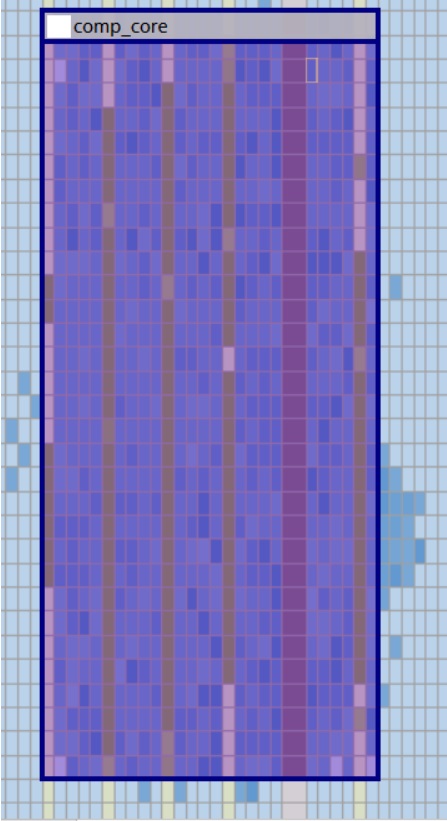}
   \caption{Tightly Constrained Placement}
   \label{fig:tight_floorplan}
\end{figure}

We then constrained the core into a rectangular bounding box with an 86\% logic utilization. The clock rate still exceeded 950 MHz. To evaluate the effects of assembling systems (multiple SIMT processors plus accelerators), we then further constrained the core into a bounding box with a 93\% utilization. This floorplan is shown in figure~\ref{fig:tight_floorplan}. Table~\ref{tab:resources} shows the resource type and distribution used by this instance. The reported logic includes unreachable ALMs inside the bounding box. Although the number of registers may appear out of proportion to the number of ALMs (which would have the effect of wasting logic, as ALMs would need to be used purely to provide registers), this is not the true case. Where possible, registers are specified without a reset, allowing the use of Agilex hyper-registers~\cite{ChromczakAgilex}. The SP has the largest concentration of registers, and for the example used here, the number of primary registers used was 763, the secondary registers 154 (these are the two additional registers in each ALM), and 420 hyper registers.

\begin{table}
\footnotesize
  \begin{center}
    \caption{SIMT Processor with Various Memory Banks and Architectures}
    \label{tab:resources}
    \begin{tabular}{|c||c|c|c|c|c|c|c||}
\hline
   \textbf{Module} & \textbf{No.} & \textbf{Sub}  & \textbf{ALMs} & \textbf{Regs} & \textbf{M20K} & \textbf{DSP} \\
      \hline\hline
      \textbf{GPGPU} & {-} & {-} & {7038} & {24534} & {99} & {32}   \\
    \hline
    \multirow{3}{*}{SP} & {16} & {-} & {371} & {1337} & {4} & {2}    \\
    \ & {-} & {Mul+Sft} & {145} & {424} & {0} & {2}   \\
    \ & {-} & {Logic} & {83} & {424} & {0} & {0}   \\
    \hline
    \ {Inst} & {1} & {-} & {275} & {651} & {3} & {0}   \\
    \hline
    \ {Shared} & {1} & {-} & {133} & {233} & {64} & {0}   \\
    \hline
    \hline
    \end{tabular}
  \end{center}
\end{table}

\subsection{Multi-Processor Designs}

We then placed 3 cores in a group, separated by a sector boundary. We ran 5-seeds of both the tightly constrained single instance and the three stamp system, with the results in   table~\ref{tab:stamps}. There was a slight clock rate hit of 3\% for the tightly constrained processor, and a further 8\% performance drop for the multi-core system.

\begin{table}
\footnotesize
  \begin{center}
    \caption{Stamping}
    \label{tab:stamps}
    \begin{tabular}{c|c|c}
    \hline
    \ & \textbf{1-Stamp} & \textbf{3-Stamp} \\
    \hline\hline
    \textbf{Best Compile} & {927 MHz} & {854 MHz} \\
    \hline
    \end{tabular}
  \end{center}
\end{table}

This occurs because many modern place and route tools optimize designs based on worst case slack~\cite{Slack_Paper} for each clock network on a given design.  When there are several instances of a design, such as in the 3-stamp case, on the same clock network it means the compiler will be simultaneously optimizing all stamps.  The worst-case slack at any point in the compile may be limited, and contained within a single stamp.  This makes it more difficult for the place and route tool to maintain performance when additional instances are added. There are a number of possibilities to mitigate this effect, and we feel that a system performance ({\em{i.e.}} a design consisting of multiple SIMT cores plus some accelerators) of 850 MHz is a reasonable target.

\section{Future Work}

In this paper, we demonstrated a repeatable and reliable SIMT processor compiling into a current FPGA at close to 1 GHz. We were also able to maintain this performance level with an location and geometry constrained compile. There are three immediate research areas that our results point to.

The latter compile was constrained at the macro level (the Quartus tool had the freedom to arrange the design across the designated region); the next step will be to explore component level constraints, such as aligning individual SPs to individual rows or regions (encompassing the minimum required number of M20Ks and DSP Blocks for that instance). Being able to control placement on a fine level will increase the density of system packing; for example, packing at the SP level will allow a sector to be filled completely, as the additional pipeline stage needed to maintain performance at the near 1 GHz level across the sector boundary (between the SPs and the shared memory) can be placed precisely where needed.  

In this work, we implemented a single processor. A multi-processor design will show how the FPGA can support high performance systems. This will encompass both packing processors together (whether continuing the approach shown in this paper, or using the results of the component level research above), and combining with a high speed interconnect fabric. While multi-processor systems are not new to FPGA, this new work will also have to maintain very high performance (possibly over 900MHz). Note that the multi-stamp example is comprised of three individual processors, and is a simpler place and route problem that this proposed investigation.

Current results are all based on logic based approaches: first maximizing the logic density, and then matching the combination of logic and embedded resources to the structure of FPGA. But routing is hardly considered; the relationship between the many 32-bit busses required by the processor and the hierarchical routing architecture of the Agilex devices needs to be evaluated. For example, the logic based shifters could not maintain 1 GHz in a larger system setting, largely because of routing distance. We only looked at the problem, and found an alternate solution, by extending the multiplier to perform both logical and arithmetic shifts. Ultimately, a routing driven placement method (or at least analysis) needs to be developed, in order to understand why some bottlenecks occur during high speed designs, and how these could be mitigated by using the FPGA differently.  

\section{Conclusions}

We have shown that high-performance soft processors - approaching the 1 GHz design limit of current FPGA devices - can be realized. We presented a parallel SIMT processor which exceeded 950 MHz clock frequency, limited only by the performance of the DSP Block. All compiles were achieved for both unconstrained and constrained designs, with only a minimal amount of synthesis or placement directives. These results are repeatable, including the placement of multiple instances of the processor onto a single device, where a speed of 850 MHz can be achieved even with the core packed to a high 93\% logic utilization.  

High performance design for the FPGA is possible if the structure of the device is considered at each step of the design process. Just increasing the pipeline depth of a circuit is not good enough, as this may impact the structure of the datapath, making it more difficult to fit.

These results clearly point the way to further work. Further exploration in constrained fitting both offer the promise of greater design density, as well as driving a better understanding of high-speed FPGA design. Developing multi-processor systems will expand the capability of the FPGA, potentially providing a processor system that can approach the performance of datapath accelerators the FPGA is known for.

\footnotesize
\bibliographystyle{IEEEtran}
\bibliography{references.bib}

\begin{thebibliography}{10}
\providecommand{\url}[1]{#1}
\csname url@samestyle\endcsname
\providecommand{\newblock}{\relax}
\providecommand{\bibinfo}[2]{#2}
\providecommand{\BIBentrySTDinterwordspacing}{\spaceskip=0pt\relax}
\providecommand{\BIBentryALTinterwordstretchfactor}{4}
\providecommand{\BIBentryALTinterwordspacing}{\spaceskip=\fontdimen2\font plus
\BIBentryALTinterwordstretchfactor\fontdimen3\font minus \fontdimen4\font\relax}
\providecommand{\BIBforeignlanguage}[2]{{%
\expandafter\ifx\csname l@#1\endcsname\relax
\typeout{** WARNING: IEEEtran.bst: No hyphenation pattern has been}%
\typeout{** loaded for the language `#1'. Using the pattern for}%
\typeout{** the default language instead.}%
\else
\language=\csname l@#1\endcsname
\fi
#2}}
\providecommand{\BIBdecl}{\relax}
\BIBdecl

\bibitem{FPL2024_proceedings}
\BIBentryALTinterwordspacing
\emph{2024 34th International Conference on Field-Programmable Logic and Applications {FPL} Turin, September 2-6, 2024}.\hskip 1em plus 0.5em minus 0.4em\relax {IEEE}, 2024. [Online]. Available: \url{https://ieeexplore.ieee.org/xpl/conhome/10705425/proceeding}
\BIBentrySTDinterwordspacing

\bibitem{Nios}
Intel, \emph{Nios II Classic Processor Reference GuideNios II Classic Processor Reference Guide}, 2016, \url{https://www.intel.com/content/www/us/en/docs/programmable/683620\\ /current/overview-67435.html}.

\bibitem{NiosV}
\emph{Nios V Processor Reference Manual}, 2022, \url{https://www.intel.com/content/www/us/en/products/details/fpga/nios-processor/v.html}.

\bibitem{Microblaze}
\emph{Microblaze Processor Reference Guide}, 2018, \url{https://docs.xilinx.com/v/u/2018.2-English/ug984-vivado-microblaze-ref}.

\bibitem{FlexGrip}
K.~Andryc, M.~Merchant, and R.~Tessier, ``Flexgrip: A {S}oft {GPGPU} for {FPGAs},'' in \emph{2013 International Conference on Field-Programmable Technology (FPT)}, 2013, pp. 230--237.

\bibitem{MIAOW}
\BIBentryALTinterwordspacing
R.~Balasubramanian, V.~Gangadhar, Z.~Guo, C.-H. Ho, C.~Joseph, J.~Menon, M.~P. Drumond, R.~Paul, S.~Prasad, P.~Valathol, and K.~Sankaralingam, ``Enabling {GPGPU} {L}ow-{L}evel {H}ardware {E}xplorations with {MIAOW}: An {O}pen-{S}ource {RTL} {I}mplementation of a {GPGPU},'' \emph{ACM Trans. Archit. Code Optim.}, vol.~12, no.~2, jun 2015. [Online]. Available: \url{https://doi.org/10.1145/2764908}
\BIBentrySTDinterwordspacing

\bibitem{FGPU}
\BIBentryALTinterwordspacing
M.~Al~Kadi, B.~Janssen, and M.~Huebner, ``{FGPU}: An {SIMT}-{A}rchitecture for {FPGA}s,'' in \emph{Proceedings of the 2016 ACM/SIGDA International Symposium on Field-Programmable Gate Arrays}, ser. FPGA '16.\hskip 1em plus 0.5em minus 0.4em\relax New York, NY, USA: Association for Computing Machinery, 2016, p. 254–263. [Online]. Available: \url{https://doi.org/10.1145/2847263.2847273}
\BIBentrySTDinterwordspacing

\bibitem{SCRATCH}
\BIBentryALTinterwordspacing
P.~Duarte, P.~Tomas, and G.~Falcao, ``{SCRATCH}: {A}n {E}nd-to-{E}nd {A}pplication-{A}ware {S}oft-{GPGPU} {A}rchitecture and {T}rimming {T}ool,'' in \emph{Proceedings of the 50th Annual IEEE/ACM International Symposium on Microarchitecture}, ser. MICRO-50 '17.\hskip 1em plus 0.5em minus 0.4em\relax New York, NY, USA: Association for Computing Machinery, 2017, p. 165–177. [Online]. Available: \url{https://doi.org/10.1145/3123939.3123953}
\BIBentrySTDinterwordspacing

\bibitem{DOGPU}
R.~Ma, J.-C. Hsu, T.~Tan, E.~Nurvitadhi, R.~Vivekanandham, A.~Dasu, M.~Langhammer, and D.~Chiou, ``{DO-GPU}: Domain {O}ptimizable soft {GPU}s,'' in \emph{2021 31st International Conference on Field-Programmable Logic and Applications (FPL)}, 2021, pp. 140--144.

\bibitem{Guppy}
A.~Al-Dujaili, F.~Deragisch, A.~Hagiescu, and W.-F. Wong, ``{G}uppy: A {GPU}-like soft-core processor,'' in \emph{2012 International Conference on Field-Programmable Technology}, 2012, pp. 57--60.

\bibitem{Kingyens}
\BIBentryALTinterwordspacing
J.~Kingyens and J.~G. Steffan, ``The {P}otential for a {GPU}-{L}ike {O}verlay {A}rchitecture for {FPGA}s,'' \emph{Int. J. Reconfigurable Comput.}, vol. 2011, pp. 514\,581:1--514\,581:15, 2011. [Online]. Available: \url{https://doi.org/10.1155/2011/514581}
\BIBentrySTDinterwordspacing

\bibitem{Vegas}
\BIBentryALTinterwordspacing
C.~H. Chou, A.~Severance, A.~D. Brant, Z.~Liu, S.~Sant, and G.~G. Lemieux, ``{VEGAS:} soft vector processor with scratchpad memory,'' in \emph{Proceedings of the {ACM/SIGDA} 19th International Symposium on Field Programmable Gate Arrays, {FPGA} 2011, Monterey, California, USA, February 27, March 1, 2011}, J.~Wawrzynek and K.~Compton, Eds.\hskip 1em plus 0.5em minus 0.4em\relax {ACM}, 2011, pp. 15--24. [Online]. Available: \url{https://doi.org/10.1145/1950413.1950420}
\BIBentrySTDinterwordspacing

\bibitem{Venice}
\BIBentryALTinterwordspacing
A.~Severance and G.~Lemieux, ``{VENICE:} {A} compact vector processor for {FPGA} applications,'' in \emph{2012 International Conference on Field-Programmable Technology, {FPT} 2012, Seoul, Korea (South), December 10-12, 2012}.\hskip 1em plus 0.5em minus 0.4em\relax {IEEE}, 2012, pp. 261--268. [Online]. Available: \url{https://doi.org/10.1109/FPT.2012.6412146}
\BIBentrySTDinterwordspacing

\bibitem{VectorBlox}
\BIBentryALTinterwordspacing
A.~Severance and G.~G.~F. Lemieux, ``Embedded supercomputing in {FPGA}s with the {VectorBlox} {MXP} {M}atrix {P}rocessor,'' in \emph{Proceedings of the International Conference on Hardware/Software Codesign and System Synthesis, {CODES+ISSS} 2013, Montreal, QC, Canada, September 29 - October 4, 2013}.\hskip 1em plus 0.5em minus 0.4em\relax {IEEE}, 2013, pp. 6:1--6:10. [Online]. Available: \url{https://doi.org/10.1109/CODES-ISSS.2013.6658993}
\BIBentrySTDinterwordspacing

\bibitem{ISFPGA_eGPU}
\BIBentryALTinterwordspacing
M.~Langhammer and G.~A. Constantinides, ``{A} {S}tatically and {D}ynamically {S}calable {S}oft {GPGPU},'' in \emph{Proceedings of the 2024 ACM/SIGDA International Symposium on Field Programmable Gate Arrays}, ser. FPGA '24.\hskip 1em plus 0.5em minus 0.4em\relax New York, NY, USA: Association for Computing Machinery, 2024, p. 165–175. [Online]. Available: \url{https://doi.org/10.1145/3626202.3637567}
\BIBentrySTDinterwordspacing

\bibitem{Nvidia_PTX}
Nvidia, \emph{PTX ISA Release 8.4}, 2024, \url{//docs.nvidia.com/cuda/pdf/ptxisa8.4.pdf}.

\bibitem{Agilex_DSP}
\emph{Intel Agilex 7 Variable Precision DSP Blocks Overview}, 2023, \url{https://www.intel.com/content/www/us/en/docs/programmable/683037/23-3/variable-precision-dsp-blocks-overview.html}.

\bibitem{Agilex}
Intel, \emph{Intel Agilex {FPGAs} and {SOCs}}, 2021, \url{https://www.intel.com/content/www/us/en/products/programmable/fpga// agilex.html}.

\bibitem{Agilex_Logic}
------, \emph{Intel Agilex® 7 Logic Array Blocks and Adaptive Logic Modules User Guide}, 2023, \url{https://www.intel.com/content/www/us/en/docs/programmable/683577// current/register.html}.

\bibitem{ChromczakAgilex}
\BIBentryALTinterwordspacing
J.~Chromczak, M.~Wheeler, C.~Chiasson, D.~How, M.~Langhammer, T.~Vanderhoek, G.~Zgheib, and I.~Ganusov, ``Architectural {E}nhancements in {I}ntel{\textregistered} agilex{\texttrademark} {FPGA}s,'' in \emph{{FPGA} '20: The 2020 {ACM/SIGDA} International Symposium on Field-Programmable Gate Arrays, Seaside, CA, USA, February 23-25, 2020}, S.~Neuendorffer and L.~Shannon, Eds.\hskip 1em plus 0.5em minus 0.4em\relax {ACM}, 2020, pp. 140--149. [Online]. Available: \url{https://doi.org/10.1145/3373087.3375308}
\BIBentrySTDinterwordspacing

\bibitem{Slack_Paper}
P.~Liao, S.~Liu, Z.~Chen, W.~Lv, Y.~Lin, and B.~Yu, ``Dreamplace 4.0: Timing-driven global placement with momentum-based net weighting,'' in \emph{2022 Design, Automation and Test in Europe Conference and Exhibition (DATE)}, 2022, pp. 939--944.

\end{thebibliography}

\end{document}